\documentclass[
 reprint,
 superscriptaddress,
 amsmath,amssymb,
 aps,
prl,
]{revtex4-2}

\usepackage{graphicx}
\usepackage{dcolumn}
\usepackage{bm}
\usepackage{hyperref}
\usepackage{siunitx}
\usepackage{xspace}
\usepackage{xcolor}
\usepackage{sidecap}
\usepackage{braket}
\usepackage[T1]{fontenc}

\graphicspath{{final_submission_PRX/}}
\let\biblio\bibliography
\let\bibsty\bibliographystyle
\renewcommand{\bibliography}[1]{\expandafter\biblio{final_submission_PRX/#1}}
\renewcommand{\bibliographystyle}[1]{\expandafter\bibsty{final_submission_PRX/#1}}

\begin{document}

\newcommand{\partitle}[1]{\subsection*{#1}}

\title{Realization of a cavity-coupled Rydberg array}

\author{Jacopo De Santis}\thanks{These authors contributed equally to this work.}
    \affiliation{Max-Planck-Institut f\"{u}r Quantenoptik, 85748 Garching, Germany}
    \affiliation{Munich Center for Quantum Science and Technology (MCQST), 80799 Munich, Germany}
    \affiliation{Fakult\"{a}t f\"{u}r Physik, Ludwig-Maximilians-Universit\"{a}t, 80799 Munich, Germany}
\author{Balázs Dura-Kovács}\thanks{These authors contributed equally to this work.}
    \affiliation{Max-Planck-Institut f\"{u}r Quantenoptik, 85748 Garching, Germany}
    \affiliation{Munich Center for Quantum Science and Technology (MCQST), 80799 Munich, Germany}
    \affiliation{Fakult\"{a}t f\"{u}r Physik, Ludwig-Maximilians-Universit\"{a}t, 80799 Munich, Germany}
\author{Mehmet Öncü}\thanks{These authors contributed equally to this work.}
    \affiliation{Max-Planck-Institut f\"{u}r Quantenoptik, 85748 Garching, Germany}
    \affiliation{Munich Center for Quantum Science and Technology (MCQST), 80799 Munich, Germany}
    \affiliation{Fakult\"{a}t f\"{u}r Physik, Ludwig-Maximilians-Universit\"{a}t, 80799 Munich, Germany}
\author{Adrien Bouscal}\thanks{These authors contributed equally to this work.}
    \affiliation{Max-Planck-Institut f\"{u}r Quantenoptik, 85748 Garching, Germany}
    \affiliation{Munich Center for Quantum Science and Technology (MCQST), 80799 Munich, Germany}
    \affiliation{Fakult\"{a}t f\"{u}r Physik, Ludwig-Maximilians-Universit\"{a}t, 80799 Munich, Germany}
\author{Dimitrios Vasileiadis}
    \affiliation{Max-Planck-Institut f\"{u}r Quantenoptik, 85748 Garching, Germany}
    \affiliation{Fakult\"{a}t f\"{u}r Physik, Ludwig-Maximilians-Universit\"{a}t, 80799 Munich, Germany}
    \affiliation{Technische Universit\"{a}t M\"{u}nchen, TUM School of Natural Sciences, 85748 Garching, Germany}
\author{Johannes Zeiher}
    \affiliation{Fakult\"{a}t f\"{u}r Physik, Ludwig-Maximilians-Universit\"{a}t, 80799 Munich, Germany}
    \affiliation{Max-Planck-Institut f\"{u}r Quantenoptik, 85748 Garching, Germany}
    \affiliation{Munich Center for Quantum Science and Technology (MCQST), 80799 Munich, Germany}

\date{\today}

\begin{abstract}
Scalable quantum computers and quantum networks require the combination of quantum processing nodes with efficient light-matter interfaces to distribute quantum information in local or long-distance quantum networks.
Neutral-atom arrays have both been coupled to Rydberg states to enable high-fidelity quantum gates in universal processing architectures, and to optical cavities to realize interfaces to photons.
However, combining these two capabilities and coupling atom arrays to highly excited Rydberg states in the mode of an optical cavity has been an outstanding challenge.
Here we present a novel cavity-coupled Rydberg array that achieves this long-standing goal.
We prepare, detect, and control individual atoms in a scalable optical tweezer array, couple them strongly to the optical mode of a high-finesse optical cavity and excite them in a controlled way to Rydberg states.
We show that high-cooperativity coupling to an optical cavity -- demonstrated via the dispersive shift of the resonance of the cavity in presence of the atoms -- and strong Rydberg interactions -- demonstrated via the collective enhancement of Rydberg coupling in the atomic array -- can be achieved in our setup at the same spatial location.
Our presented experimental platform opens the path to several new directions, including the realization of quantum network nodes, quantum simulation of long-range interacting, open quantum systems and photonic-state engineering leveraging high-fidelity Rydberg control.
\end{abstract}

\maketitle

\section{Introduction}
The realization of efficient and high-quality interfaces between emitters and single photons is a central goal in quantum science~\cite{Chang2014}, and an essential requirement for distributed quantum computing, metrology, or quantum communication, and ultimately a scalable quantum internet~\cite{Kimble2008,Wehner2018,Covey2023}.
Optical cavities are a leading approach to achieving strong coupling between single atoms and single photons in the strong-coupling or high-cooperativity regime of cavity quantum electrodynamics (cQED)~\cite{Reiserer2015}.
Single- or few-atom cavity-couplings have been leveraged to realize elementary atom-based building blocks of quantum networks, including fast non-destructive detection of atoms~\cite{Bochmann2010, Volz2011, Deist2022}, the creation of entanglement between atoms~\cite{Welte2018,Daiss2021,Grinkemeyer2025} or the realization of photonic entangled states~\cite{Thomas2022,Thomas2024}.
Very recently, the capabilities of atomic cQED systems have been expanded by placing arrays of single atoms trapped in optical tweezers inside single-mode cavities~\cite{Deist2022a,Deist2022,Hartung2024,Ho2025,Wang2025,Hu2025,Peters2025} or cavity arrays~\cite{Shaw2026,Soper2026}.
In parallel, arrays of atoms coupled to highly excited Rydberg states have rapidly progressed, with advances in quantum computing, quantum simulation and quantum metrology~\cite{Browaeys2020,Kaufman2021}.
In particular, quantum computing based on neutral-atom arrays has become a central focus in the field, with breakthroughs in the realization of high-fidelity gates in different species and qubit implementations~\cite{Evered2023,Muniz2025,Peper2025,Tao2026,Senoo2026}, excellent scalability of the array sizes~\cite{Gyger2024,Norcia2024,Manetsch2025,Chiu2025,Zhu2026} and first demonstrations of logical qubits and circuits~\cite{Bluvstein2024,Reichardt2025,Zhang2026a,Rines2025,Bluvstein2026}. 

Complementing the excellent control over individual qubits in Rydberg arrays with an efficient atom-light interface featuring strong coupling between single atoms and single photons opens the path towards remote entanglement between nodes with computing capabilities, which is a cornerstone of scalable neutral-atom-based quantum network nodes~\cite{Reiserer2022a,Huie2021,Covey2023,Zhang2025b}, smaller-scale logical quantum memories~\cite{Sinclair2025}, and ultimately distributed quantum computing architectures~\cite{Monroe2014,Li2024,Sunami2025}.
%
\begin{figure*}
    \centering
    \includegraphics{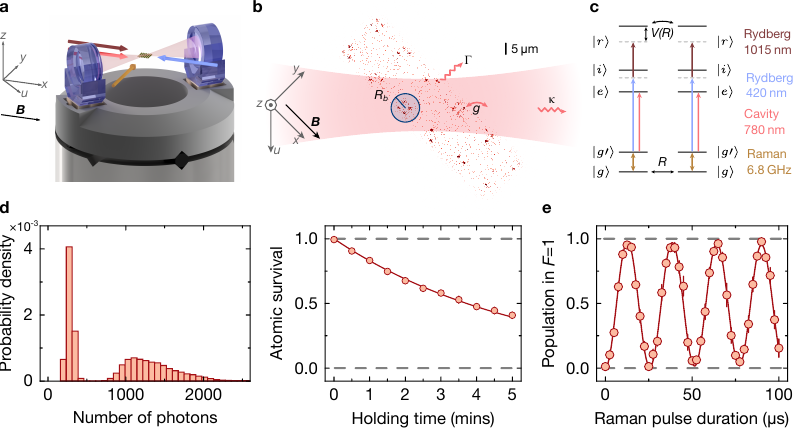}
    \caption{\textbf{Experimental setup and array generation.}
 \textbf{a} Close-up of the optical cavity including the electric-field shielding platform, and atomic array. The counter-propagating dark red (blue) beams along the $x$-axis represent the \SI{1015}{\nano\meter} (\SI{420}{\nano\meter}) components of the Rydberg excitation. The light red arrow represents the polarizer beam pumping to the state $\ket{F=2, m_F=-2}$. The orange arrow represents the Raman beam. The light red shading between the mirrors denotes the cavity mode (not to scale). Along the vertical axis, the optical tweezers are projected through an out-of-vacuum objective (not shown) above the cavity. Atoms from the 2D MOT enter the science region from below, passing through the opening in the science platform, loading a 3D MOT at the center of the cavity.  \textbf{b}~Array of Rydberg-blockaded ensembles of $2\times2$ tweezers inside the cavity mode volume (red, not to scale), stochastically loaded with $^{87}\mathrm{Rb}$ atoms. The magnetic quantization field is aligned with the $x$-axis. The Rydberg blockade disk is indicated in blue. The single-atom coupling to cavity mode $g$, the cavity decay rate $\kappa$, and the free-space atomic decay rate $\Gamma$ are indicated. \textbf{c} Level diagram for two atoms, including the Raman, Rydberg, and cavity-coupling transitions. $\ket{g^\prime} = \ket{5\mathrm{S}_{1/2}, F=2, m_F = -2}, \ket{g} = \ket{5\mathrm{S}_{1/2}, F=1, m_F = -1}, \ket{e} = \ket{5\mathrm{P}_{3/2}, F=3}, \ket{i} = \ket{6\mathrm{P}_{3/2}, F=3}, \ket{r} = \ket{53\mathrm{S}_{1/2}}$. For a distance $R<R_b$, the Rydberg interaction energy shift $V$ leads to Rydberg blockade. \textbf{d} Loading individual atoms in the optical traps, we perform high-fidelity ($99.988(3)\%$) and high-survival ($99.88(2)\%$) imaging with bimodal count histograms. Measurement of the loss dynamics of tweezer-trapped atoms yields a lifetime of \SI{322(3)}{\second}. \textbf{e} Starting from atoms optically pumped to the state $\ket{F=2, m_F = -2}$ using circularly polarized light, we demonstrate high-fidelity manipulations of the internal states of $^{87}\mathrm{Rb}$ via a Raman beam detuned from the $\mathrm{D}_1$ line. All errorbars represent the standard error of the mean (s.e.m.).}

    \label{fig:1}
\end{figure*}
Promising efforts towards the realization of such a hybrid platform include nano-photonic cavities~\cite{Dordevic2021,Menon2024}.
However, the strong electric fields in the vicinity of such devices have to be managed in order to generate Rydberg-based entanglement, which is limited to regions distant from the device~\cite{Ocola2024}.
In general, integrating high-fidelity Rydberg coupling with optical cavities poses challenges due to the presence of dielectric surfaces, e.g., in near-planar~\cite{Daiss2021,Thomas2022,Hartung2024,Thomas2024} or fiber-based cavities~\cite{Hunger2010,Grinkemeyer2025}, or due to the electric field created by the required piezoelectric transducers, which are used to stabilize the length of optical cavities and ensure strong coupling to the atoms.
While exciting progress has been made in photonic-state-engineering by ensembles coupled both to the modes of optical cavities and Rydberg states~\cite{Ningyuan2016, Stolz2022,Vaneecloo2022}, these experiments typically do not operate in the single-atom high-cooperativity cQED regime and are lacking single-atom control.
Up to now, it has been an outstanding challenge in the field to couple scalable arrays of single atoms to Rydberg states in or even near the mode of optical cavities.

Here, we overcome several experimental challenges and combine a tweezer array with individual atoms strongly coupled to the optical mode of a high-finesse optical cavity with Rydberg excitation.
We prepare, detect, and control individual atoms in a scalable optical tweezer array with up to $48$ traps with high fidelities and couple them strongly to the mode of an optical cavity.
We characterize the atom-cavity coupling via the dispersive atom-photon interaction when detuning the cavity from atomic resonance, and show non-destructive state-selective detection using the cavity.
Going beyond previous work, we controllably excite Rydberg states at the location of the cavity mode.
In particular, we demonstrate that the transition frequency to Rydberg states is only minimally perturbed by the electric fields due to the cavity.
Finally, we leverage the arbitrary configurability of the optical tweezers to prepare ensembles of up to four individual atoms in the regime of Rydberg blockade.
Optically coupling these blockaded ensembles to a high-lying Rydberg state, we observe a square-root scaling of the Rabi frequency with atom number, a characteristic signature of entangled W-states expected in this regime~\cite{Dudin2012,Zeiher2015}.
%

%
\section{Experimental setup}
Our experiments are performed in a new experimental machine based on $^{87}\mathrm{Rb}$ atoms, optimized for combining a tweezer array, a near-concentric optical cavity and Rydberg excitations; see Fig.~\ref{fig:1}a.
Our tweezer arrays at a wavelength of \SI{1015}{\nano\meter} are generated using a spatial light modulator, which enables arbitrary tweezer configurations via phase-only modulation~\cite{Nogrette2014}.
The tweezers are focused in the atomic plane using a microscope objective with a numerical aperture (NA) of $0.6$, resulting in arrays of up to $49$ near diffraction-limited optical tweezers at the center of our optical cavity with a trap depth of approximately \SI{1.2}{\milli\kelvin}.
Fluorescence from the atoms is collected through the same objective and then delivered to a qCMOS camera, yielding a high-signal-to-noise image of individual atoms in the tweezer array for an exposure time of \SI{80}{\milli\second}; see Fig.~\ref{fig:1}b.
By analyzing three subsequent images~\cite{Manetsch2025}, we find initial loading probabilities of $52.0(2)\%$,  imaging fidelities of $99.988(3)\%$ and single-atom survival probabilities as high as $99.88(2)\%$, compatible with other results achieved in alkali atoms; see Fig.~\ref{fig:1}d. 
The high-finesse cavity is built in a near-concentric configuration, assembled from two mirrors with radius of curvature $R=\SI{10.0(1)}{\milli\meter}$ each, and with a mirror separation of $\sim \SI{20}{\milli\meter}$.
The cavity is one-sided with input (output) mirror transmission of \SI{2}{ppm} (\SI{30}{ppm}).
This configuration maximizes optical access while simultaneously supporting small optical modes at the cavity center and hence large single-atom cooperativity $C = 4g^2/\kappa\Gamma$, where $g$ denotes the vacuum Rabi coupling of the atom to the cavity mode, $\kappa$ the decay rate of the photons out of the cavity, and $\Gamma$ the decay rate of atoms due to photons emitted into free space~\cite{Tanji2011,Periwal2021,Deist2022a,Deist2022}.
In addition to the cavity coupling, we can also controllably excite the atoms to Rydberg states on a two-photon transition via the intermediate $6\mathrm{P}_{3/2}$-state and at interatomic distances where Rydberg blockade is expected; see Fig.~\ref{fig:1}c.
In order to ensure mechanical decoupling of the cavity from vibrations of the experimental table and the vacuum chamber, we mount the two mirrors on top of a mechanical vibration isolation tower; see Fig.~\ref{fig:1}a.
The combination of stainless steel rings and an ultra-high vacuum compatible fluoroelastomer results in a strong low-passing character of the entire structure.
The top part of the tower is manufactured from titanium, which features a low outgassing rate and thus enables long vacuum-limited lifetimes.
In fact, we verify that we can hold the atoms in our tweezers under continuous cooling conditions for \SI{322(3)}{\second} (see Fig.~\ref{fig:1}d), despite the vicinity of the glued cavity mirror assemblies.
At the same time, the metallic titanium platform shields electric fields generated by the piezos used to stabilize the cavity length.

In order to demonstrate high-fidelity manipulation of the internal atomic states, we first optically pump the atoms to the state $\ket{F=2, m_F = -2}$ using a $\sigma^-$-polarized beam aligned with the quantization axis ($x$-axis) and stabilized to the $F=2 - F^\prime=2$ transition in $^{87}\mathrm{Rb}$.
Subsequently, we drive the $\ket{F=1, m_F=-1}$ to $\ket{F=2, m_F=-2}$ transition using a Raman beam aligned with the $y$-axis, with an intermediate-state detuning of $\Delta_{\mathrm{R}} = 2\pi\times \SI{100}{\giga\hertz}$ from the D1 line.
We extract a SPAM-corrected $\pi$-pulse fidelity for this transition of 99.2(3)$\%$ from the measured Rabi oscillation at the $\pi$-pulse time.
The Raman system uses a combination of a chirped Bragg grating with an electro-optical modulator to generate efficient amplitude modulation~\cite{Levine2022}.
We perform state-selective detection by removal of atoms in $F=2$ to observe Rabi oscillations and thus arbitrary initial-state control at the single-qubit level when combined with phase-control on the Raman beam pair; see Fig.~\ref{fig:1}e.
We also use the same Raman beam with a retroreflection mirror in order to drive motional sidebands of the atoms, allowing us to realize Raman sideband cooling down to average motional occupation of $\bar{n}=0.62(15)$ in one dimension.
We note that while Raman cooling is possible in principle in our setup, the measurements presented were achieved without this additional cooling step.
%
%
\begin{figure}
    \centering
    \includegraphics{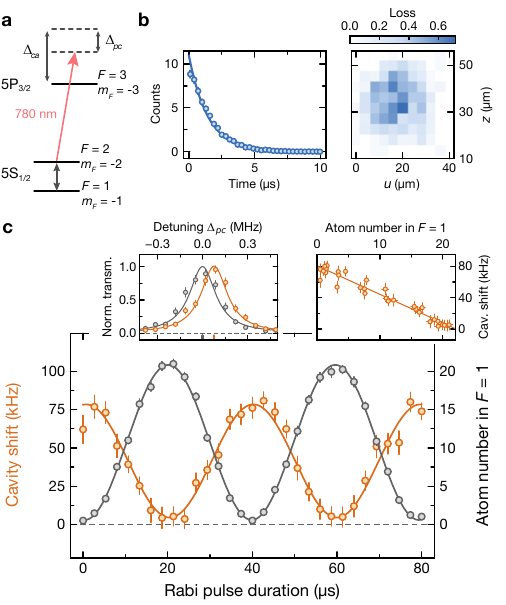}
    \caption{\textbf{Cavity geometry and atom-cavity interaction.} \textbf{a} Relevant levels to probe the cavity spectrum. The probe beam, resonant with the cavity, is detuned from the $\mathrm{D}_2$-cycling transition by $\Delta_{ca}$ and the probe is varied around that point by $\Delta_{pc}$. Atoms in $F = 1$ act as spectators not coupled to the cavity. \textbf{b}  Left panel: Photon storage time in the cavity. An exponential fit yields a decay time of $t_d = \SI{1.44(2)}{\micro \second}$, corresponding to a cavity FWHM of $\kappa = 2\pi\times \SI{110(2)} {\kilo\hertz}$. Right panel: Tomographic atom loss measurement while scanning the array position in the $z$-direction and averaging along the cavity direction. We obtain a image scan of the cavity mode, from which we extract cavity waists of $w_u =\SI{20(1)}{\micro\meter}$ and $w_z =\SI{21(1)}{\micro\meter}$. \textbf{c} Characterization of the cavity through dispersive shift measurements. The main panel shows the dispersive cavity shift after driving the system from $\ket{F=2, m_F = -2}$ to $\ket{F=1, m_F = -1}$ using a microwave Rabi pulse with variable duration (orange). The sinusoidal variation reflects the number of atoms in $F=2$ coupled to the cavity. The state-selective cavity measurement is in good agreement with a destructive state-selective fluorescence measurement of the atoms performed in the same sequence after pushing all atoms in $F=2$. Left inset: Lorentzian cavity lineshapes with (orange) and without atoms (grey) together with Lorentzian fits used to extract the cavity shift in the main panel. Orange (grey) ticks indicate the position of the cavity resonance with (without) atoms. Right inset: Atom number-resolved shift extracted from two curves shown in the main panel. We find the expected linear relation between atom number and cavity shift. All errorbars for population measurements represent the standard error of the mean (s.e.m.). For the cavity shift measurements, the error represents the 1-sigma confidence interval for the resonance positions extracted from the fits.}
    \label{fig:2}
\end{figure}
\begin{figure*}
    \centering
    \includegraphics{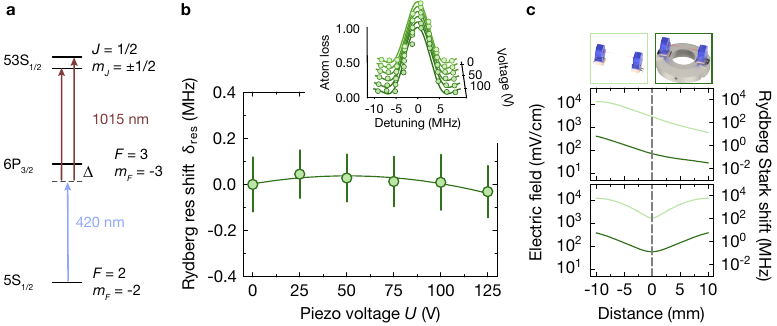}
    \caption{\textbf{Electric field shielding.}
    \textbf{a} Rydberg excitation scheme. We couple the Rydberg state $53\mathrm{S}_{1/2}$ starting from the ground state $5\mathrm{S}_{1/2}$ via the intermediate state $6\mathrm{P}_{3/2}$ with an intermediate-state detuning of $\Delta = 2\pi\times \SI{2}{\giga\hertz}$ via lasers at \SI{420}{\nano\meter} and \SI{1015}{\nano\meter}. \textbf{b} We measure the Rydberg resonance position $\delta_\mathrm{res}$ as a function of the applied voltage to the piezos, $U$. With both piezos driven symmetrically, we find only a weak shift that agrees well with our ab-initio modeling of the cavity assembly including the electric field shield when including a small offset electric field at zero piezo voltage (solid line). The inset shows the resonance scan at each piezo voltage from which we extract the shift from a fit and recording the position of the peak. The error represents the 1-sigma confidence interval for the resonance positions extracted from the fits. \textbf{c} Comparison of the electric field at the position of the atoms in a configuration where the mirrors sit on exposed piezos (light green) with our configuration where they are buried in a titanium platform (dark green). Top: Case where only one piezo is driven to \SI{100}{\volt}. Bottom: Symmetric case where both piezos are driven. The weak dependence observed in \textbf{b} is hence a direct consequence of our shielding of the piezo. We find that a suppression of the electric field by more than one order of magnitude is achieved, which corresponds to close to three orders of magnitude suppression of the resonance shifts (right axis).
    }
    \label{fig:3}
\end{figure*}
\section{Cavity coupling}
After demonstrating the generation of tweezer arrays as well as high-fidelity preparation and manipulation of atoms, we proceed by showing that they strongly couple to the mode of the optical cavity.
We first characterize the cavity without loading atoms using a \SI{780}{\nano\meter} probe beam coupled into the cavity and detected in transmission using a single-photon counting module (SPCM).
We find a free-spectral range of $\mathrm{FSR} = \SI{7.65(7)}{\giga\hertz}$, yielding a cavity length of 
$L_{\mathrm{exp}} = \SI{19.6(2)}{\milli\meter}$ consistent with our design.
To obtain an initial estimate of the cQED parameters of our setup, we measure the photon storage time in the cavity to be $t_d = \SI{1.44(2)}{\micro \second}$, which is equivalent to the cavity full width at half maximum (FWHM) linewidth $\kappa = 2\pi\times \SI{110(2)}{\kilo\hertz}$; see Fig.~\ref{fig:2}b.
Combined with the FSR, we find a cavity finesse of $\mathcal{F} = 7.0\,(1) \times 10^4$, in good agreement with an independent ringdown measurement obtained by sweeping the probe rapidly across the cavity resonance.
To extract the mode shape of the cavity mode, we prepare a stochastically loaded tweezer array with $4\times12$ traps in the $x-y$-plane, spaced by $\SI{4}{\micro\meter}$ orthogonal to the cavity axis and $\SI{12}{\micro\meter}$ along the cavity axis.
We initially increase the probe power coupled into the cavity to a level where the atoms within the mode scatter photons off-resonantly on the cycling transition $F=2 - F^\prime = 3$ in $^{87}\mathrm{Rb}$; see Fig.~\ref{fig:2}a.
We lower the tweezer depth from the imaging configuration to \SI{0.3}{\milli K}, such that this off-resonant scattering heats the atoms in the tweezer array and leads to atom loss at the position of the optical mode.
Throughout, the cavity is kept at a fixed detuning $\Delta_{ca} = 2\pi \times\SI{101(1)}{\mega\hertz}$ with respect to the atomic transition by locking it to a stabilized laser at \SI{840}{\nano\meter}.
Scanning the position of the array in the $z$-direction and averaging along the cavity axis, the loss feature reveals a tomographic image of the cavity mode (Fig.~\ref{fig:2}b) with extracted waists $w_u =\SI{20(1)}{\micro\meter}$ and $w_z =\SI{21(1)}{\micro\meter}$ in the $x-y$ plane and $z$-direction, respectively.
Combining these measurements, we extract an expected maximal cooperativity on the cycling transition of $C_{\mathrm{th}} = 6\mathcal{F}\lambda^2/(\pi^3w_uw_z) \approx20(1)$.

After completing the alignment of the array with the cavity mode, we reduce the probe power coupled into the cavity.
In this configuration, the atoms prepared in $\ket{F=2, m_F = -2}$ couple dispersively to the cavity resonance at minimal atom loss, leading to an expected atom-number dependent shift of the cavity resonance given by $\delta_N~=~NC(\Gamma\kappa/4\Delta_{ca}) = N(g^2/\Delta_{ca})$~\cite{Tanji2011}.
The cavity shift is measured by scanning the probe beam across the cavity resonance and comparing the resonance positions with and without the atoms present in the cavity; see Fig.~\ref{fig:2}c left inset.
Here, a single trace is obtained by stepping the probe through different $\Delta_{pc}$ and recording the counts on the SPCM for an integration time of \SI{100}{\micro \second}.
To obtain an experimental estimate of the cavity cooperativity, we precisely control the number of atoms coupled to the cavity mode using a microwave drive on the $\ket{F=2, m_F = -2}$ to $\ket{F=1, m_F = -1}$ transition; see Fig.~\ref{fig:2}c.
The cavity shift exhibits a clean sinusoidal variation as a function of the Rabi drive duration, directly reflecting the effective number of atoms in $F = 2$ coupled to the cavity and thus establishing a non-destructive state-selective atom number probe.
We furthermore find excellent correlation with a destructive measurement of the atoms in $F=1$ in the same sequence after pushing atoms in $F=2$ and taking a fluorescence image; see Fig.~\ref{fig:2} c, right inset.

Extracting the cooperativity from these measurements requires the effective number of atoms $N_\mathrm{eff}$ coupled to the cavity mode.
This number is obtained as $N_\mathrm{eff} =\alpha N$, with the correction factor $\alpha = 0.23(1)$ determined from the precise tomographic measurement of the mode and the array parameters in combination with the total atom number in the array, $N$, measured in fluorescence.
The correction factor includes the reduction of the coupling due to contributions from the Gaussian transverse mode shape ($0.47(1)$, calculated from the data in Fig.~\ref{fig:2}b) as well as the averaging over the $\cos^2$-axial standing wave of the cavity mode ($0.5$)~\cite{Yan2023}.
%
%
Combining the correction factor with the precisely determined cavity shift, we obtain an effective shift per atom of $\bar{\delta}_1 = \delta_N/N_\mathrm{eff} = \delta_N/(\alpha N) =  2\pi\times \SI{15.8(6)}{\kilo\hertz}$.
This yields an estimate for the single-atom cavity coupling of $g = 2\pi\times \SI{1.26(2)}{\mega\hertz}$.
Combined with the cavity linewidth $\kappa$ extracted from the photon storage time and the known decay rate $\Gamma$ of $^{87}\mathrm{Rb}$, we determine a lower bound for the cooperativity of our system of $C = 4g^2/\kappa\Gamma = 9.6(4)$.
Our cavity thus operates deep in the single-atom high-cooperativity cQED regime.
We attribute the discrepancy with the expected cooperativity $C_\mathrm{th}$ mostly to reduction of the coupling for our magnetic field configuration, cavity polarization and alignment of the array, and leave their optimization to future work.
%
\section{Rydberg resonance}
Exciting Rydberg atoms in the cavity poses a substantial challenge due to the vicinity of the dielectric cavity mirror substrates as well as the piezoelectric transducers required to control the length of the cavity.
In our near-concentric cavity geometry, effects coming from the mirror surfaces are expected to be small due to the large mirror separation.
However, voltages of up to \SI{125}{\volt} may be applied at the piezos at a distance of approximately \SI{10}{\milli\meter} from the location of the tweezer array and the cavity mode.
Rydberg states can couple to this electric field via their strong DC-polarizability, which results in detrimental, fluctuating resonances and long-term drifts at a scale of up to \SI{1}{\giga\hertz}.
Conversely, exciting Rydberg atoms enables precise probing of the local electric field environment at the location of the atoms~\cite{Osterwalder1999, Sedlacek2012}.
To characterize the effectiveness of our titanium platform to shield electric fields generated by the piezos, we couple to the state $53\mathrm{S}_{1/2}$ starting from the $5\mathrm{S}_{1/2}$ ground state.
We employ an off-resonant two-photon excitation from the intermediate $6\mathrm{P}_{3/2}$ state involving a blue photon at \SI{420}{\nano\meter} on the lower leg and a near-infrared photon at \SI{1015}{\nano\meter} on the upper leg; see Fig.~\ref{fig:3}a.
We note that, at our tweezer wavelength, the Rydberg states are antitrapped.
To mitigate trap-induced resonance asymmetries due to this trapping mismatch, we rapidly turn off the tweezers during the spectroscopy pulse and apply approximate $\pi$-pulses while scanning the two-photon detuning; see inset of Fig.~\ref{fig:3}b.
Tracking the position of the resonance as function of the applied voltage to the piezo, we find only small total shifts on the order of \SI{100}{\kilo\hertz} peak to peak for voltages up to \SI{125}{\volt}, in line with the expectation obtained from a 3D finite-element simulation of the electric field configuration at the location of the atoms performed with COMSOL on our manufacturing CAD model.
This strong suppression of the electric field is further illustrated in Fig.~\ref{fig:3}c by finite-element electrostatics simulations comparing the field as a function of the distance from the piezo in the cases of shielded and unshielded piezos.
We find that our simple design of burying the piezo in the metallic titanium platform leads to a suppression of the electric field at the position of atoms by more than one order of magnitude, resulting in a reduction of the expected resonance shifts from the \SI{100}{\mega\hertz}-range to approximately \SI{100}{\kilo\hertz}.
Furthermore, we note that it is unlikely that applying voltages as high as \SI{125}{\volt} will be required to keep the cavity at the working point.
In fact, we have confirmed that long term drifts of the cavity resonance can be compensated by controlled heating of the cavity mounts, allowing the piezos to always work in the same limited voltage range.
We anticipate that the small modulations required on top of such a large, thermally regulated cavity length, will render fluctuations of the Rydberg resonance completely irrelevant.
%

\section{Collective Rabi oscillations}
\begin{figure}
    \centering
    \includegraphics{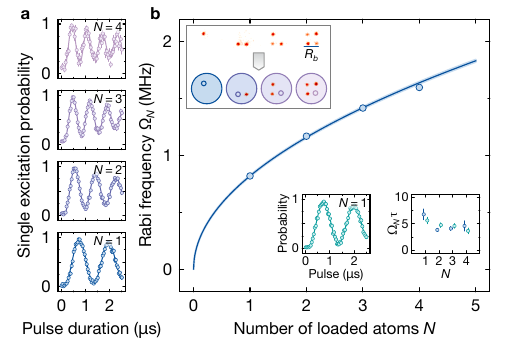}
    \caption{\textbf{Collective Rydberg coupling.}
    \textbf{a} Preparing a tweezer array with groups of four tweezers within a Rydberg blockade radius $R_b=\SI{5.8}{\micro\meter}$, indicated by the radius of the circle in the inset in \textbf{b}, we measure the probability of finding maximally one excitation in the system after suddenly turning on the Rydberg coupling light (blue points). When postselecting on the initially prepared number of atoms, $N$, we find that the Rabi frequency increases with atom number, as extracted from a cosinusoidal fit of the form $A(1-e^{-(t-t_0)/\tilde{\tau}}\cos(\tilde{\Omega}(t-t_0) + \phi))$. Errorbars represent the standard error of the mean.
    \textbf{b} We find quantitative agreement with the expected dependence of the extracted Rabi frequency (blue points) with a prediction based on scaling the extracted Rabi frequency $\Omega_{\mathrm{Ryd}}$ for $N=1$ with the square-root of the number of atoms within a blockade radius, $\Omega_N = \Omega_{\mathrm{Ryd}}\sqrt{N}$ (solid line). The left panel in the inset shows the Rabi oscillations in the presence of a full-amplitude piezo scan (cyan datapoints), together with a fit (solid line). The right panel shows the extracted ``quality factor'' $\Omega_N\tau$ without (with) scanning the piezos in blue (cyan). We find no difference between the two cases within the errorbars, which represent the 1-sigma confidence intervals extracted from the fits.}
    \label{fig:4}
\end{figure}
With stable Rydberg resonances established, we finally demonstrate the controlled excitation of Rydberg states in an interacting regime at the location of the cavity mode.
Specifically, we adapt the tweezer pattern to feature groups of atoms separated by approximately \SI{3.2}{\micro\meter}, within the expected blockade radius of $R_b = \SI{5.8}{\micro\meter}$ for the Rydberg state $53\mathrm{S}_{1/2}$ at our expected Rabi frequency $\Omega_{\mathrm{Ryd}}\simeq2\pi \times \SI{1}{\mega\hertz}$.
When suddenly switching on the Rydberg coupling under these conditions, we observe coherent Rabi oscillations between the ground and Rydberg states, which are not trapped and thus lost from our tweezers before detection.
We group the data according to the atom number obtained from a first image taken before the Rydberg excitation.
Limiting the maximum number of lost atoms to one in a second image after the dynamics, we extract the probability of creating a single Rydberg excitation as a function of the initial atom number; see Fig.~\ref{fig:4}a.
Double Rydberg excitation due to imperfect blockade or atom loss in the tweezers due to heating lead to a small fraction of $7\%$ of the events where two or more atoms are lost.
We note that the oscillation onset delay at $t=0$ comes from the finite rise time of the switching AOM.
Fitting a damped oscillation to the data for $N=1$ atom per group, we extract $\Omega_{\mathrm{Ryd}} = 2\pi\times \SI{0.820(4)}{\mega\hertz}$, with a dephasing rate $\Gamma_{\mathrm{Ryd}} = 2\pi\times \SI{0.12(2)}{\mega\hertz}$.
The residual dephasing is likely due to a combination of inhomogeneity of the coupling due to excitation beam size and finite atom temperature. 
Furthermore, we find that the Rabi frequency increases with the initial atom number $N$ in a four-tweezer group with a characteristic dependence of the coupling strength, $\Omega_N = \Omega_{\mathrm{Ryd}}\sqrt{N}$, signaling the coupling to the multi-partite entangled W-state~\cite{Dudin2012,Zeiher2015}; see Fig.~\ref{fig:4}b.
To verify the minimal impact of the electric field also in this Rabi oscillation measurement, we perform the same sequence while scanning the cavity piezos between $\SI{20}{\volt}$ and $\SI{100}{\volt}$ with a triangle ramp at a frequency of \SI{42}{\hertz}.
Extracting the dephasing time of the Rabi oscillations $\tau$, we find that the ``quality factor'' $\Omega_N \tau$, i.e. the number of coherent oscillations per decay time is, within errorbar, indistinguishable between the scanned and unscanned cases despite the large scanning amplitude.

These measurements were intentionally performed at lower Rydberg Rabi frequency to increase the blockade radius.
When operating at our maximally achievable Rydberg Rabi frequency of $\SI{4.37(2)}{\mega\hertz}$, we obtain a SPAM-corrected $\pi$-pulse fidelity of $\SI{98.0(4)}{\%}$.
From a Ramsey measurement we extract a coherence time $T_2^*=\SI{1.8(9)}{\micro\second}$, within errorbars independent of whether the piezo is driven or not.
%

\section{Conclusion}
In summary, we have presented a new experimental setup that combines single tweezer-trapped atoms with a high-cooperativity optical cavity and highly excited Rydberg states.
Our results clearly demonstrate the compatibility of these capabilities, which will be enhanced by several improvements to be implemented in the near future.
%
%
%
In particular, we are working on a phase-noise reduction system for our Rydberg excitation beam at \SI{1015}{\nano\meter}, which is currently derived from an extended-cavity diode laser locked to an ultralow expansion glass cavity.
We expect that the suppression of phase noise in the MHz-regime~\cite{DeLeseleuc2018, Jiang2023, Denecker2025} will open a straightforward path towards high-fidelity two-qubit gate operations on the ground-state qubits.
Already with the currently achieved parameters, our setup opens the route to a number of interesting directions, including the realization of fast cavity-assisted non-destructive readout~\cite{Deist2022} and cyclic error correction codes prepared using high-fidelity Rydberg coupling~\cite{Bluvstein2024,Bluvstein2026}.
Fast, non-destructive readout combined with high-fidelity preparation of entangled states could also be used to realize one-way-quantum computing~\cite{Raussendorf2001}. 
The non-local connectivity provided by the cavity opens the route towards novel two-qubit gate schemes~\cite{Borregaard2015,Chen2015,Ramette2022,Ramette2025} that could overcome the time-overhead of shuttling qubits over large distances within a node by asynchronous entanglement distribution.
Furthermore, leveraging the high-fidelity entangling gates of atom arrays coupled to the optical cavity, we anticipate applications in photonic state engineering, where atomic entanglement is repeatedly swapped to photonic states~\cite{Economou2010, Thomas2024, Rubies-Bigorda2025}, thus paving the way for novel quantum network~\cite{Covey2023,Zhang2025b} and distributed quantum computing~\cite{Sinclair2025} protocols.
Finally, our system opens a new direction in analog and mixed analog-digital quantum simulations where strong Rydberg interactions compete with long-range cavity interactions and dissipation.
Such a setting features a rich and experimentally largely unexplored phase diagram~\cite{Gelhausen2016,Rohn2020}, novel out-of-equilibrium dynamics~\cite{Zhang2013,Mikheev2025,Hosseinabadi2026}, as well as novel routes to engineer interesting entangled states within the cavity~\cite{Mann2025}.

\begin{acknowledgments}
We acknowledge helpful discussions with Tracy Northup, Dan Stamper-Kurn, Gerhard Rempe, and Darrick Chang and their teams.
We are grateful to Pranav Kulkarni, Sebastian Ruffert, Miriam Keim, Mullai Sampangi, and Arda Özkut for early contributions to the experimental setup. We thank Philip Thomas for contributions to the new cavity characterization. We thank Maximilian Ammenwerth for feedback on the manuscript.
We acknowledge funding by the Deutsche Forschungsgemeinschaft (DFG, German Research Foundation) under Germany's Excellence Strategy--EXC-2111--390814868 (MCQST), from the Munich Quantum Valley initiative as part of the High-Tech Agenda Plus of the Bavarian State Government, and from the BMFTR through the program “Quantum technologies---from basic research to market” (SNAQC, Grant No. 13N16265, 13N17337).
We also acknowledge funding through JST-DFG 2024: Japanese-German Joint Call for Proposals on “Quantum Technologies” (Japan-JST-DFG-ASPIRE 2024) under DFG Grant No. 554561799.
J.D.S, B.D.K, and M.Ö acknowledge funding from the International Max Planck Research School (IMPRS) for Quantum Science and Technology. A.B acknowledges support through a fellowship from the Alexander von Humboldt Foundation.
\end{acknowledgments}

\section{Data availability}

The data underlying the experiments are available via Edmond~\cite{DeSantis2026data}.


%

\clearpage


\clearpage

\end{document}